\documentstyle[aps,multicol,epsf]{revtex}
\newcommand{\be}{\begin{equation}}
\newcommand{\ee}{\end{equation}}
\newcommand{\amu}{a_{\mu}}
\newcommand{\asi}{a_{\sigma}}
\begin{document}
\draft
\title{Quantum Dot and Hole Formation in Sputter Erosion \\}
\author{B. Kahng$^{1,2}$, H. Jeong$^1$, and A.-L. Barab\'asi$^1$ \\}
\address{
$^1$ Department of Physics, University of Notre Dame, 
Notre Dame, IN 46556 \\
$^2$ Department of Physics and Center for Advanced Materials 
and Devices, Konkuk University, Seoul 143-701, Korea \\}
\date{\today}
\maketitle
\thispagestyle{empty}
\begin{abstract}
Recently it was experimentally demonstrated that sputtering 
under normal incidence leads to the formation of spatially 
ordered uniform nanoscale islands or holes. Here we show 
that these nanostructures have inherently nonlinear origin, 
first appearing when the nonlinear terms start to dominate 
the surface dynamics. Depending on the sign of the nonlinear 
terms, determined by the shape of the collision 
cascade, the surface can develop regular islands or holes with 
identical dynamical features, and while the size of these 
nanostructures is independent of flux and temperature, it 
can be modified by tuning the ion energy.    
\end{abstract}
\begin{multicols}{2}
\narrowtext

The fabrication and physical properties of quantum dots (QDs) 
are topics of high current interest due to their 
applications in optical devices and as potential construction 
blocks of novel computer architectures\cite{review}.
However, despite the high interest in the subject, 
the methods available for the fabrication of  
such dots are rather limited. 
Lithographic techniques, while undergoing rapid improvements 
in resolution\cite{reed}, still cannot produce small and dense 
enough dots necessary for device applications. 
While much research has focused on self-assembled QD formation \cite{sad}, 
that generates dots through the unique combination of 
strain and growth kinetics, these techniques offered 
relatively uniform islands only for a few material 
combinations and still have to find their way into devices. 
Consequently, there is continued high demand for alternative 
methods that would allow low cost and efficient mass fabrication 
of QDs. In the light of these technological 
driving forces, the recent demonstration that 
sputter erosion can lead to uniform nanoscale 
islands, that exhibit quantum confinement, will undoubtly 
capture the interest of the scientific community \cite{sputtering}. 
Ion beam sputtering has long been a leading candidate 
for surface patterning. While ripple formation on 
sputter eroded surfaces has been observed already in 
the 70s\cite{vasiliu}, in the last decade much work has been 
devoted to understand both the experimental and theoretical 
aspects of this fascinating self-organized phenomena. However, 
most experiments have focused on off-normal 
incidence, that, by breaking the symmetry along the surface, 
leads to anisotropic structures, such as ripples. 
While theoretically it was expected that 
under normal incidence the ripples 
should be replaced by some periodic cellular structures, 
such surface features have not been observed experimentally. 
Recently, two groups have obtained simultaneous advances 
in this direction. Facsko $et$ $al.$, investigating low-energy 
normal incident Ar$^+$ sputtering of GaSb (100) 
surfaces\cite{sputtering}, observed that as erosion proceeds, 
nanoscale islands appear on the surface, that are remarkably 
well ordered, and have a uniform size distribution. 
On the other hand, recent experiments of Ar$^{+}$ 
sputtering of Cu(110), and Ne$^{+}$ sputtering of Ag (001) 
under normal incidence lead to relatively 
uniform depressions or holes \cite{italy}. 
These experiments, while represent significant technological 
breakthroughs, raise a number of questions regarding the 
mechanisms responsible for the formation of these nanostructures. 
While it is tempting to interpret these structures as periodic 
perturbations expected by the linear theory of sputtering 
\cite{harper}, a careful analysis of the experimental results indicates 
that such approach is less than satisfactory.
To mention only a few discrepancies, Facsko $et$ $al.$~reported 
that the obtained island size is independent of temperature, 
while, according to the linear theory
their size should decrease exponentially with $T$ \cite{harper}.
Also, the linear theory would predict at best a cellular structure 
displaying a square lattice, in contrast with the hexagonal ordering 
observed in experiments \cite{sputtering}. \\

In this paper we present the first detailed theory addressing
the formation of sputter-induced QDs and holes.  
We demonstrate that these structures are the result of 
inherently nonlinear phenomena, and thus they cannot be accounted for in 
the context of the linear theory. We find that regular 
nanostructures first appear as the nonlinear terms become 
relevant, allowing us to predict the characteristic time, $\tau$, 
necessary for their formation, and calculate the dependence 
of $\tau$ on the physical parameters characterizing the ion 
bombardment process. Furthermore, we show that the QDs and holes, 
observed in different experiments, are governed by the same 
physical phenomena, the difference between them coming from 
the shape of the ion cascade characterizing the interaction 
of the bombarding ions with the substrate. 
Finally, we predict that the size of these islands and 
holes, while independent of flux and temperature, 
depends on the ion energy. \\

A particularly successful description of the morphological 
evolution of sputter eroded surfaces has been proposed by 
Bradley and Harper (BH)\cite{harper}, based on Sigmund's 
theory of sputtering\cite{sigmund}, predicting that the 
height $h(x,y,t)$ of the eroded surface is described by 
the linear equation 
\begin{equation}
{\partial_t h} = \nu_x \partial_x^2 h +\nu_y \partial_y^2 h
-K\partial^4 h,
\end{equation}
where $\nu_x$ and $\nu_y$ are effective surface tensions,
generated by the erosion process and $K$ is the surface 
diffusion constant induced by thermal diffusion.
The balance of the unstable erosion term ($-|\nu| \partial^2 h$)
and the surface diffusion term ($-K\partial^4 h$) acting to 
smooth the surface, generates ripples with wavelength
\begin{equation}
\ell_i=2\pi \sqrt{2K/|\nu_i|},
\end{equation}
where $i$ refers to the direction ($x$ or $y$) along
which $\nu_i$ ($\nu_x$ or $\nu_y$) is
the largest. While successful in predicting the ripple 
wavelength and orientation\cite{koponen}, the linear 
theory fails to explain a number of experimental features, 
such as the saturation of the ripple amplitude\cite{wittmaack,aziz10,vajo},
or the appearance of kinetic roughening\cite{eklund,yang}. 
To address these questions it has been proposed \cite{cuerno} 
that these shortcomings can be cured by the inclusion of nonlinear 
terms and noise, derived from the Sigmund's theory of sputtering. 
Consequently, Eq.~(1) has to be replaced by the nonlinear equation,
\begin{eqnarray}
{\partial_t h} &=& \nu_x \partial_x^2 h +\nu_y \partial_y^2 h
-D_x\partial_x^4 h -D_y \partial_y^4 h-D_{xy}\partial_x^2
\partial_y^2 h \nonumber \\
& &+{\lambda_x \over 2}(\partial_x h)^2 + {\lambda_y \over 2}
(\partial_y h)^2 + \eta(x,y,t),
\end{eqnarray}
where 
$\lambda_x$ and $\lambda_y$ describe the tilt-dependent erosion rate, 
depending on flux $f$ and the penetration depth $a$, 
and $\eta(x,y,t)$ is an uncorrelated
white noise with zero mean, mimicking the randomness resulting 
from the stochastic nature of ion arrival to the surface 
\cite{cuerno,maxim1}. 
Furthermore, the sputtering process always generates 
ion induced effective diffusion, that together with thermal 
diffusion, are incorporated in $D_x$, $D_y$, and $D_{xy}$
\cite{maxim1}.
Under normal incidence, the coefficients in (3) are isotropic, 
given by \cite{maxim1,maxim2} 
\begin{equation}
\nu \equiv \nu_x=\nu_y =-fa \asi^2/2 \amu^2, 
\end{equation}
\begin{equation}
D\equiv  D_x=D_y=fa^3\asi^2/8\amu^4,
\end{equation} 
\begin{equation}
\lambda \equiv \lambda_x=\lambda_y =(f/2\amu^2)(\asi^2-\asi^4-\amu^2), 
\end{equation}
where $\amu=a/\mu$ and $\asi=a/\sigma$ and $\mu$ and $\sigma$, 
defined in Fig.~1a, characterize the shape of the collision 
cascade of the bombarding ion.\\

To investigate the origin of the QDs and the dynamics of 
QD formation under normal incident ion sputtering, 
we integrated numerically the continuum equation (3), 
using standard discretization techniques\cite{numerical}, 
and isotropic coefficients as expected for normal incidence. 
We choose a temporal increment $\Delta t=0.01$ and impose 
periodic boundary conditions $h(x,y,t)=h(x+L,y,t)=h(x,y+L,t)$ 
where $L \times L$ is the size of the substrate. 
To improve the uniformity of the QDs, 
the numerical simulations were carried out without noise  
($\eta=0$), using instead a random initial 
surface configuration. However, we repeated the simulations 
for the noisy case as well, finding that the size uniformity 
of QDs is enhanced as the noise amplitude decreases. \\

Our main result, presenting the morphology of the ion sputtered surface 
at three different stages of their time evolution, is shown in Fig.~2. 
Let us first concentrate on the $\lambda > 0$ case (upper panels in 
Fig.~2). In the early stages of the sputtering process the surface 
is dominated by small, wavy perturbations (Fig.~2a) generated 
by the interplay between the ion induced instability and surface 
relaxation. However, since the system is isotropic in the $(x,y)$ 
plane, these ripple precursors are oriented randomly, generating 
short wormlike morphologies on 
the surface. After some characteristic time, $\tau$, these structures 
turn into isolated but closely packed islands, reminiscent of the QDs 
reported experimentally (Fig.~2b). 
Note that upon a closer inspection one can observe 
the emergence of hexagonal order in the island positions. 
As the sputtering proceeds, while the islands do not disappear, 
the supporting surface develops a rough profile, destroying the overall 
uniformity of the islands (Fig.~2c). A similar scenario is observed 
for $\lambda <0$, the only difference being that now the islands 
are replaced by holes (Figs.~2d-f), reminiscent of the morphologies 
observed experimentally by Rusponi $et$ $al.$\cite{italy}. 
The first conclusion we can draw from these results is that 
the development of QDs and holes is governed by the same underlying 
physical phenomena, the only difference being that for QDs we 
have $\lambda >0$, and for holes $\lambda <0$. 
Indeed, this morphological change is expected from the 
nonlinear continuum theory, 
Eq.~(3) being symmetric under the simultaneous transformation 
$\lambda \rightarrow -\lambda$ and $h \rightarrow -h$, 
indicating that changing the sign of $\lambda$ does not 
affect the dynamics of the surface evolution, but  
simply turns the islands into mirrored holes. 
Since, according to (6) the sign of $\lambda$ is determined 
only by the relative magnitude of $a_{\sigma}$ and $a_{\mu}$, 
whether islands or holes appear is determined by the shape of 
the collision cascade, shown in Fig.~1a. Consequently, 
using (6) we can draw a phase diagram in terms of the reduced 
penetration depths $a_{\sigma}$ and $a_{\mu}$ that separates 
the regions displaying QDs versus holes (Fig.~1b).   
These results also indicate that the QDs and holes are 
inherently nonlinear objects, that can be explained only by the nonlinear 
theory, since, should the linear terms be responsible for their 
formation, the surface morphology 
should not depend on the sign of $\lambda$ (indeed, Eq.~(1) 
has a full $h \rightarrow -h$ symmetry). 
In the following, we investigate the dynamics of QD and hole formation, 
providing further proof of their common nonlinear origin. \\

The crossover behavior from the linear to the nonlinear regimes
can be monitored through the surface width, 
$W^2 (L,t) \equiv {1 \over L^2}\sum_{x,y} h^2(x,y,t)-{\bar h}^2$.
As Fig.~3 shows, this quantity exhibits a sharp transition 
at a characteristic time $\tau$: for $t < \tau$, the width $W$ 
increases exponentially as $W \sim \exp(\nu t/\ell^2)$,
while for $t > \tau$, $W$ still increases but at a considerably 
smaller rate than an exponential\cite{spark}. 
The crossover time is given by $\tau \sim \sigma^2/fa$
in terms of the experimental parameters\cite{spark}, indicating 
its dependence on the flux and ion beam energy. Correlating 
these results with the observed surface morphologies, we find 
that the QDs first appear at $t\approx \tau$. 
Indeed, in Fig.~3 we marked with arrows the time when the 
morphologies in Fig.~2a-c were recorded, indicating 
that no QDs exist before $\tau$ (Fig.~2a), but they are fully developed 
at $\tau$ (Fig.~2b), and their uniformity rapidly vanishes after $\tau$ 
(Fig.~2c).\\

Based on the results presented in Figs.~2 and 3, the following scenario 
emerges for QD and hole formation.
In the early stages of the erosion process the linear theory 
correctly describes the surface evolution, and thus we observe 
the cell structure predicted by the BH theory. 
However, as time increases, the nonlinear terms 
$turn$ $on$ breaking the up-down symmetry of the surface. 
Such delayed effect of the nonlinear terms is a well known 
feature of the class of surface evolution equations which 
(3) belongs to \cite{book}.
The sign of $\lambda$ determines whether QDs or holes 
form, these structures appearing at a characteristic 
time $\tau$, that depends on the flux and the ion energy, 
and their characteristic size is given by the ripple 
period $\ell$ (Eq.~(2)). 
In turn, the sign of $\lambda$ is determined only by the shape 
of the collision cascade (Fig.~1).
As time increases beyond $\tau$, the nonlinear terms 
lead to kinetic roughening of the surface at large length 
scales\cite{cuerno,book}, and while the QDs and holes 
do not disappear, the substrate 
on which they exist becomes rough, destroying the overall 
island/hole uniformity and ordering.
Indeed, as Fig.~3b shows, the island height distribution has a larger width
after $\tau$ than right at $\tau$.
Thus, ordered and uniform QDs and holes can be obtained only 
in the early stages of the nonlinear regime, more precisely,
at the crossover between the linear and nonlinear regimes 
emerging at the characteristic time $\tau$. \\

Experimentally the crossover time $\tau$ can be estimated 
through monitoring the evolution of the ripple amplitude 
using $in$ $situ$ spectroscopic techniques\cite{aziz10}. 
Our results indicate that in order to obtain uniform islands 
one needs to stop the sputtering process just when the amplitude
of the eroded surface saturates: before saturation the QDs 
have not developed yet, while after saturation their uniformity 
disappears. Since the mechanism generating the QDs is 
the ion induced instability, balanced by ion induced diffusion, 
their typical separation is given by the most unstable wavelength, 
$\ell=2\pi \sqrt{2D/\nu}$.
The simulations indicate that the islands (or holes) first 
appear as small surface perturbations. However, 
as their amplitude increases, around $\tau$ they get in 
contact with each other, and their growth saturates. 
Thus their final diameter coincides with $\ell$. 
Indeed, we determined the diameter of the QDs in simulations 
for $\nu=-0.6169$, $D=2$ and $\lambda=1$, observing 
approximately 16 QDs aligned along the side of the system 
of the size $256 \times 256$, in agreement with the 
prediction $L/\ell\approx 16$.
Using (4) and (5), we find that $\ell=\sqrt{2} \pi \mu$, 
i.e., from the $average$ $separation$ of the islands one can 
$determine$ $the$ $size$ $of$ $the$ $horizontal$ $width$ of the 
collision cascade (Fig.~1a). Furthermore, since typically we have 
$\mu \sim a \sim \epsilon^{2m}$, when $\epsilon$ is the ion energy 
and $m$ is an constant that weakly depend on $\epsilon$ 
($m=1/2$ for $\epsilon \approx 10 \sim 100$ keV)\cite{sigmund}, 
we predict that one can tune the size of the QDs by changing 
the ion energy $\epsilon$, while the size is independent of the flux 
and the temperature.\\     

Finally, the conclusion $\ell \sim \mu$ holds only when ion 
induced effective diffusion \cite{maxim1} is the main relaxation 
mechanism. While the temperature independent QD size reported 
by Facsko $et$ $al.$ \cite{sputtering} indicates that for GaSb 
this is the main relaxation mechanism, at high temperatures or 
in other systems thermal diffusion could be more relevant 
\cite{maxim1,maxim2,mac}.
Note, however, that the scenario presented above for QD and hole 
formation, is not conditional on ion induced diffusion. 
Should thermal diffusion be the dominating relaxation mechanism 
(certainly true in any system at high temperatures) it will 
change only our prediction for $\ell$ and $\tau$. 
In this case we expect $\ell \sim 2\pi \sqrt{2D_0 \exp(-E/2kT)/\nu}$, 
thus the QD size will depend exponentially on $T$, 
and we also have $\ell \sim \epsilon^{-1/2}$, and $\ell \sim f^{-1/2}$. 
For the crossover time we obtain $\tau \sim D_0^2 \exp(-2E/kT)/\nu$, 
$\tau \sim \epsilon^{-1}$, and $\tau \sim f^{-1}$. 
However, the phase diagram (Fig.~1b) 
and the expected dynamical evolution (Figs.~2 and 3) 
will not be sensitive to the nature of the relaxation mechanism.\\   

In conclusion, we demonstrated that the nonlinear 
theory can fully explain the recently observed quantum dots 
and holes generated by sputter erosion. We showed that these 
nanostructures appear only when the nonlinear terms 
become effective, stabilizing the surface amplitude, and 
that the difference in the islands and holes comes in 
the sign of the nonlinear terms. Furthermore, we are 
able to predict the dependence of the characteristic time 
for QD formation, and find that their saturation size is 
given by the size of the collision cascade. 
We believe that these results will lead to a better 
understanding of the formation and evolution of these fascinating 
nanostructures, thus they will guide further 
experiments aiming to better control these systems.\\  

This work was supported by ONR, NSF, and the Korean Research 
Foundation (Grant No. 99-041-D00150).\\


\begin{figure}
\centerline{\epsfxsize=8.3cm \epsfbox{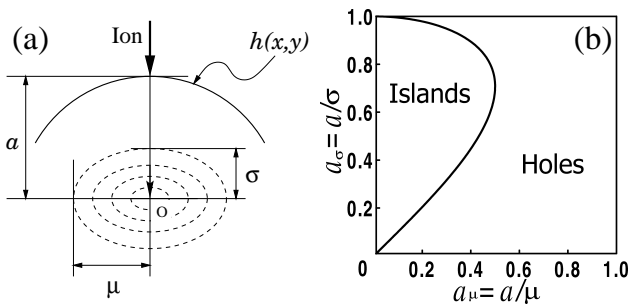}}
\caption{(a) The characteristics of the collision 
cascade generated by an ion. According to Sigmund's theory 
of sputtering [8], an ion penetrates the 
surface until a peneration depth $a$, where it radiates its energy 
out, such that the kinetic energy generated damage, decays 
exponentially with the distance from $O$. 
For most systems the energy distribution is anisotropic, being 
characterized by the characteristic decay lengthscales 
$\sigma$ and $\mu$ along or perpendicular to the 
ion direction, respectively.
(b) Phase diagram showing the parameter regimes corresponding 
to island (QD, $\lambda >0$) and hole ($\lambda <0$) formation.} 
\label{fig1}
\end{figure}
\begin{figure}
\centerline{\epsfxsize=8.3cm \epsfbox{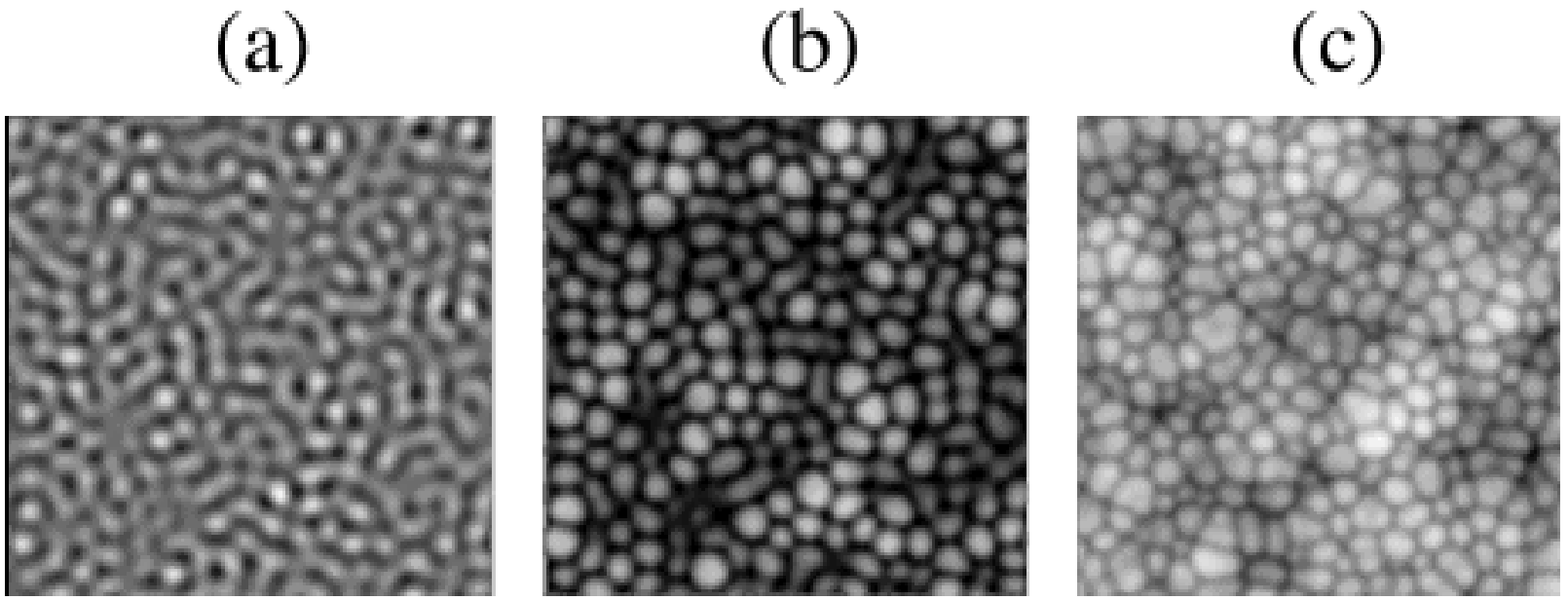}}
\centerline{\epsfxsize=8.3cm \epsfbox{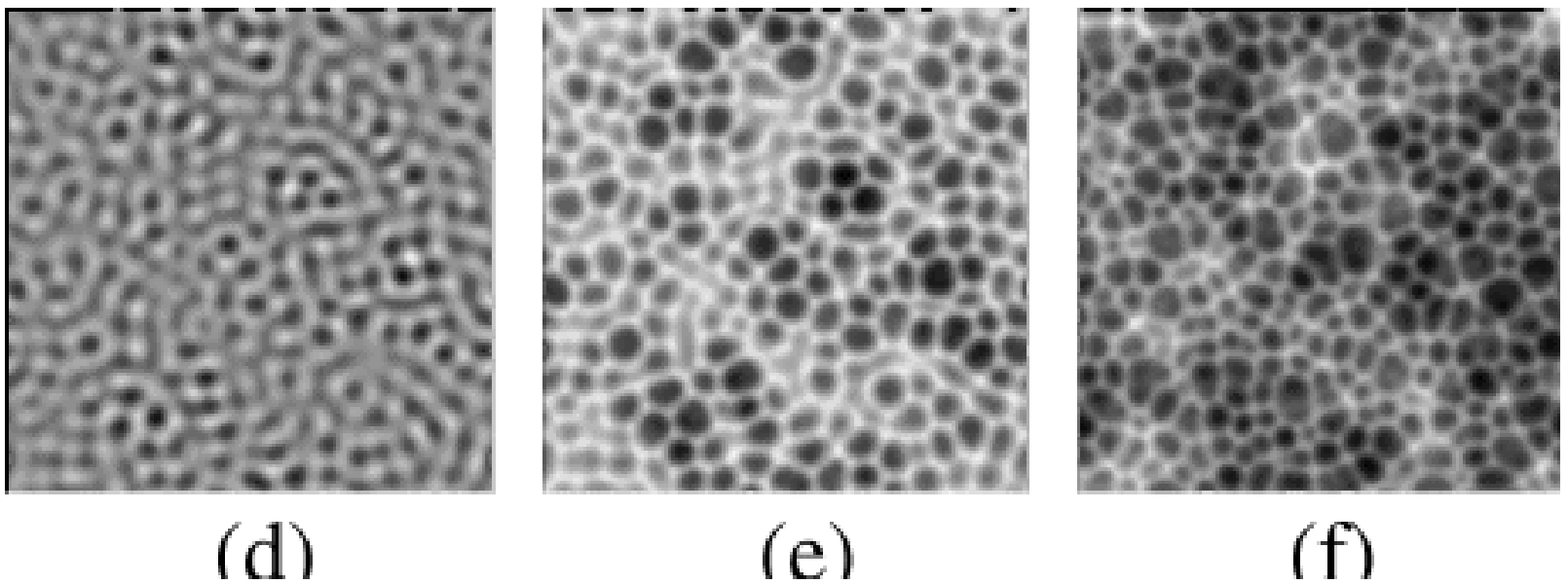}}
\caption{(a)-(c) Surface morphologies predicted by Eq.~(3), 
for $\lambda=1$ at different stages of surface evolution. 
The pictures correspond to (a) $t=$4.0, (b) 5.8, (c) 8.0$\times 
10^4$. (d)-(f) The same as in (a)-(c), but for $\lambda=-1$. 
In all cases we used $\nu=0.6169$, $D=2$, and system 
size $256 \times 256$.}
\label{fig2}
\end{figure}
\begin{figure}
\centerline{\epsfxsize=8.3cm \epsfbox{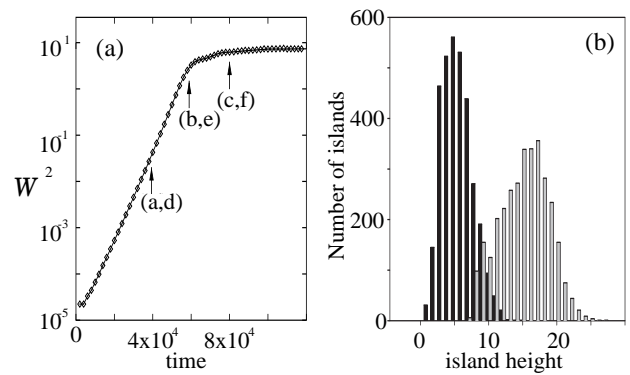}}
\caption{(a) Time evolution of the surface width $W^2$ 
for the parameters $\nu=-0.6169$, $D=2$, $\lambda=1$ or $-1$, 
and system size $256 \times 256$.
The crossover time is estimated to be $\tau \approx 5.8 \times 10^4$. 
The arrows denote the moments when the surface snapshots shown in Fig.~2 
were recorded. (b) Island height distribution right at the
crossover time $t=\tau$ (black columns), and at a later time, t=8.0$\times 10^4$ (gray columns), indicating
that the width of the distribution is smaller at the crossover time, and broadens as we go beyond $\tau$.}
\label{fig3}
\end{figure}

\end{multicols}

\begin{thebibliography}{99} 
\bibitem{review} For a review, for example, see L. Jacak, P. Hawrylak, 
and A. Wojs, {\it Quantum dots} (Springer-Verlag, Berlin, 1998).  
\bibitem{reed} T.I. Kamins and R.S. Williams, Appl. Phys. Lett. 
{\bf 71}, 1201 (1997). 
\bibitem{sad} For a review, see V.A. Shchukin and D. Bimberg, 
Rev. Mod. Phys. {\bf 71}, 1125 (1999). 
\bibitem{sputtering} S. Facsko, T. Dekorsy, C. Koerdt, C. Trappe, 
H. Kurz, A. Vogt, and H.L. Hartnagel, {\rm Science} 
{\bf 285}, 1551 (1999).
\bibitem{vasiliu} F. Vasiliu, I.A. Teodorescu, and F. Glodeanu, 
J. Mater. Sci. {\bf 10,} 399 (1975). 
\bibitem{italy} S. Rusponi, C. Boragno, and U. Valbusa,
{\rm Phys. Rev. Lett.} {\bf 78,} 2795 (1997); {\it ibid} {\bf 78,} 
4184 (1998); S. Rusponi, G. Costantini, F. B. de Mongeot, 
C. Boragno, and U. Valbusa, Appl. Phys. Lett. {\bf 75,} 3318 (1999).
\bibitem{harper} R.M. Bradley and J.M.E. Harper, {\rm J. Vac.
Sci. Technol. A} {\bf 6}, 2390 (1988).
\bibitem{sigmund} P. Sigmund, {\rm Phys. Rev.} {\bf 184,} 383
(1969).
\bibitem{koponen} I. Koponen, M. Hautala, and O.-P. Sievaenen, {\rm Phys.
Rev. Lett.} {\bf 78,} 2612 (1997).
\bibitem{wittmaack} K. Wittmaack,
{\rm J. Vac. Sci. Technol. A} {\bf 8}, 2246 (1990).
\bibitem{aziz10} J. Erlebacher, M.J. Aziz, E. Chason,
M.B. Sinclair, and J.A. Floro, (preprint).
\bibitem{vajo} J.J. Vajo, R.E. Doty, and E.-H. Cirlin,
{\rm J. Vac. Sci. Technol. A} {\bf 6,} 76 (1988).
\bibitem{rot_exp} S. Rusponi, G. Costantini, C. Boragno, and U. Valbusa,
{\rm Phys. Rev. Lett.} {\bf 81,} 2735 (1998).
\bibitem{eklund} E.A. Eklund $et$ $al.$, 
{\rm Phys. Rev. Lett.} {\bf 67}, 1759 (1991).
\bibitem{yang} H.-N. Yang, G.-C. Wang, and T.-M. Lu,
{\rm Phys. Rev. B} {\bf 50}, 7635 (1994).
\bibitem{cuerno} R. Cuerno and A.-L. Barab\'asi,
{\rm Phys. Rev. Lett.} {\bf 74}, 4746 (1995).
\bibitem{maxim1} M.A. Makeev and A.-L. Barab\'asi, Appl. Phys.
Lett. {\bf 71}, 2800 (1997).
\bibitem{maxim2} M. Makeev, R. Cuerno, and A.-L. Barab\'asi (preprint).
\bibitem{numerical} W.H. Press, B.P. Flannery, S.A. Teukolsky,
and W.T. Vetterling, {\it Numerical Recipes} (Cambridge Univ.
Press, Cambridge, 1986).
\bibitem{spark} S. Park, B. Kahng, H. Jeong, and A.-L. Barab\'asi, 
Phys. Rev. Lett. {\bf 83}, 3486 (1999).
\bibitem{book} A.-L. Barab\'asi and H.E. Stanley, $Fractal$ 
$Concepts$ $in$ $Surface$ $Growth$ (Cambridge University Press, 
Cambridge, 1995).
\bibitem{mac}
S. W. MacLaren, J. E. Baker, N.L. Finnegan, and C. M. Loxton, J. Vac. Sci.
Technol. A {\bf 10}, 468 (1992); C. C. Umbach, R. L. Headrick, 
B. H. Cooper, J.M. Balkely, E. Chason, Bull. Am. Phys. Soc. {\bf 44,} 
706 (1999).
\end{thebibliography}
\end{document}